\documentclass{article}

\usepackage{PRIMEarxiv}

\usepackage[utf8]{inputenc} % allow utf-8 input
\usepackage[T1]{fontenc}    % use 8-bit T1 fonts
\usepackage{hyperref}       % hyperlinks
\usepackage{url}            % simple URL typesetting
\usepackage{booktabs}       % professional-quality tables
\usepackage{amsfonts}       % blackboard math symbols
\usepackage{nicefrac}       % compact symbols for 1/2, etc.
\usepackage{microtype}      % microtypography
\usepackage{lipsum}
\usepackage{fancyhdr}       % header
\usepackage{graphicx}       % graphics
\graphicspath{{media/}}     % organize your images and other figures under media/ folder
\usepackage{enumerate}
\usepackage{enumitem}
\usepackage{listings}

\usepackage[natbib, bibencoding=utf8, citestyle=numeric, bibstyle=ieee, maxbibnames=999, maxcitenames=2, mincitenames=1, sortcites]{biblatex}
\bibliography{references}

\usepackage{xcolor}
\usepackage[frozencache,cachedir=minted]{minted} % for publishing
\usepackage{minted}
\usepackage{tabularx}

\usepackage{xspace}
\newcommand{\newtextitcmd}[2]{%
  \expandafter\newcommand\csname #1\endcsname{{\textit{#2}}\xspace}
}

\newtextitcmd{au}{autrainer}
\newtextitcmd{os}{openSMILE}
\newtextitcmd{ds}{DeepSpectrum}
\newtextitcmd{pann}{PANN}
\newtextitcmd{hf}{Hugging Face}
\newtextitcmd{wav}{wav2vec2.0}
\newtextitcmd{hub}{HuBERT}
\newtextitcmd{astmodel}{AST}
\newtextitcmd{mlflow}{MLflow}
\newtextitcmd{tensorboard}{TensorBoard}
\newtextitcmd{hydra}{Hydra}
\newtextitcmd{optuna}{Optuna}

\usepackage{todonotes}
\usepackage[acronym, shortcuts, nohypertypes={acronym}]{glossaries}
\newacronym{AI}{AI}{artificial intelligence}
\newacronym{ML}{ML}{machine learning}
\newacronym{DL}{DL}{deep learning}
\newacronym{AAC}{AAC}{Automated Audio Captioning}
\newacronym{ASC}{ASC}{Acoustic Scene Classification}
\newacronym{SED}{SED}{Sound Event Detection}
\newacronym{ASR}{ASR}{Automatic Speech Recognition}
\newacronym{CLI}{CLI}{command line interface}
\newacronym{MLOps}{MLOps}{machine learning operations}
\newacronym{CNN}{CNN}{Convolutional Neural Network}
\newacronym{RNN}{RNN}{Recurrent Neural Network}
\newacronym{DNN}{DNN}{Deep Neural Network}
\newacronym{MFCC}{MFCC}{Mel Frequency Cepstral Coefficients}
\newacronym{MLP}{MLP}{Multi-layer Perceptron}
\newacronym{SVM}{SVM}{Support Vector Machine}
\newacronym{LM}{LM}{Lanugage Model}

\usepackage[capitalize]{cleveref}
\crefname{lstlisting}{listing}{listings}
\Crefname{lstlisting}{Listing}{Listings}

\usepackage{pifont}

\newcommand\autitle{autrainer: A Modular and Extensible Deep Learning Toolkit for Computer Audition Tasks}
\title{
    \autitle
    \thanks{
        \textit{\underline{Citation}}: \textbf{Publication}
    } 
}
\author{
  \textbf{Simon Rampp}$^{1}$, \textbf{Andreas Triantafyllopoulos}$^{1,3,4}$,  \textbf{Manuel Milling}$^{1,3,4}$, \textbf{Björn W. Schuller}$^{1,2,3,4}$ \\
  $^1$CHI -- Chair of Health Informatics, Technical University of Munich, Munich, Germany \\
  $^2$GLAM -- Group on Language, Audio, \& Music, Imperial College, London, UK\\
  $^3$MCML -- Munich Center for Machine Learning, Munich, Germany \\
  $^4$MDSI -- Munich Data Science Institute, Munich, Germany\\
  \texttt{\{simon.rampp;andreas.triantafyllopoulos;manuel.milling;schuller\}@tum.de}
}
\begin{document}
\maketitle

% reference paper:
%https://www.jmlr.org/papers/volume24/22-1021/22-1021.pdf

\begin{abstract}
This work introduces the key operating principles for \au, our new deep learning training framework for computer audition tasks.
\au is a PyTorch-based toolkit that allows for rapid, reproducible, and easily extensible training on a variety of different computer audition tasks.
Concretely, \au offers low-code training and supports a wide range of neural networks as well as preprocessing routines.
In this work, we present an overview of its inner workings and key capabilities.\\
Code: \url{https://github.com/autrainer/autrainer}\\
Documentation: \url{https://autrainer.github.io/autrainer/}\\
Models: \url{https://huggingface.co/autrainer}\\
Code License: MIT
\end{abstract}

\keywords{Computer Audition \and Reproducibility \and PyTorch \and Neural Networks \and Deep Learning \and Artificial Intelligence}

\section{Introduction}
Reproducibility, code quality, and development speed constitute the `impossible trinity' of contemporary experimental \ac{AI} research.
Of the three, the first has attracted the most attention in recent literature~\citep{kapoor2022leakage}, as reproducibility of findings is a cornerstone of science.
However, the impact of the other two should not be underestimated.
Development speed allows the quick iteration of ideas -- a necessary prerequisite in experimental sciences and a prominent feature of \ac{AI} research, as asserted by ``The Bitter Lesson'' of R. Sutton~\citep{sutton2019the}.
Similarly, code quality can be the key differentiating factor when it comes to ``standing on the shoulders of giants'', as shaky foundations can lead to a spectacular collapse.

This is why \emph{toolkits} that are easy-to-use and provide pre-baked reproducibility are critical for the proliferation and adaptation of new ideas.
The not-so-recent renaissance of \ac{DL} has been largely driven by the creation of such toolkits.
\textsc{TensorFlow}~\footnote{\url{https://www.tensorflow.org/}}, \textsc{PyTorch}~\footnote{\url{https://pytorch.org/}}, and \textsc{transformers}~\footnote{\url{https://huggingface.co/docs/transformers}} are many among numerous other toolkits that have `democratised' the use and development of \ac{DL} algorithms.
Yet, despite the fact that several of those toolkits feature some support for the audio community, their initial development with other modalities in mind (primarily images or text) has resulted in a lineage of design choices that makes them less suited for audio.

In the present work, we introduce \au as a remedy to this state of affairs.
It is an `audio-first' automated low-code training framework, offering an easily configurable interface for training, evaluating, and applying numerous audio \ac{DL} models for classification and regression tasks.
\au can be used via a \ac{CLI} and Python \ac{CLI} wrapper, which share the same functionality.
In addition, we release a set of models that have been trained with \au and can be used off-the-shelf with its inference interface.
These cover a wide gamut of computer audition tasks, aiming to showcase the flexibility of our pipeline and aid with the democratisation of training and applying \ac{DL} models for audio.

\section{Related work}
The development of domain-specific toolkits has played an essential role in advancing \ac{DL} research across various modalities, including computer audition.
While numerous toolkits and frameworks address specific aspects of the research workflow, -- such as feature extraction, data augmentation, or model training -- few offer comprehensive, end-to-end solutions.

\emph{Feature extraction} toolkits such as openSMILE~\citep{eyben2010opensmile} focus primarily on hand-crafted audio descriptors targeting speech and music analysis. Librosa\footnote{\url{https://github.com/qiuqiangkong/torchlibrosa}}~\citep{mcfee2015librosa} offers widely-used methods for generating standard audio representations like log-Mel spectrograms or \acp{MFCC}.
Audiomentations\footnote{\url{https://github.com/iver56/audiomentations}} and similar libraries\footnote{\url{https://github.com/asteroid-team/torch-audiomentations}}\footnote{\url{https://github.com/audeering/auglib}}\footnote{\url{https://github.com/facebookresearch/AugLy}} provide waveform- and spectrogram-level augmentations for improving model robustness.

Beyond that, several toolkits target \emph{model training}.
auDEEP~\citep{freitag2018audeep} generates features from spectrograms using unsupervised training methods to train \acp{SVM} and \ac{MLP} classifiers.
DeepSpectrum(Lite)~\citep{amiriparian2017snore,amiriparian2022deepspectrumlite} translates audio spectrograms into visual representations for training image models, while End2You~\citep{tzirakis2018end} supports training \acp{CNN} and \acp{RNN} with audio and spectrogram inputs.

Among \emph{end-to-end} toolkits, nkululeko~\citep{burkhardt2022nkululeko} offers feature extraction, augmentation, classical \ac{ML} and \ac{DL} training, and post-analysis of features.
SpeechBrain~\citep{ravanelli2021speechbrain} is tailored for speech processing and conversational AI, emphasising flexible configuration and transformer architectures.
ESPNet~\citep{watanabe2018espnet} offers numerous \ac{DNN} training recipes, primarily targeting \ac{ASR} and language modelling tasks.

\section{\au}
\label{sec:principles}
In this section, we describe the key operating principles of \au.
We begin with its configuration management, followed by the data pipeline, training, and inference mode.
As previously stated, the user can interact with \au using its builtin \ac{CLI} and Python \ac{CLI} wrapper.

\subsection{Hydra configurations}
\au configures its various components using \textit{Hydra}\footnote{\url{https://hydra.cc/}} -- an open-source framework for scalable configuration management based on \textit{YAML} files.
This allows for a low-code approach where the user can specify their key hyperparameters in a \textit{YAML} file.
New functionality can be incorporated by specifying paths to local Python files and classes or functions implemented therein.
For instance, this can be used to designate a new model architecture that has been locally trained by the user or implement a custom, local dataset.
As an example, \cref{conf:entry} illustrates an \au configuration, defining a computation graph where a network of the PANN~\citep{kong2020panns} family (CNN10) is trained on an \ac{ASC} (DCASE2016Task1-16k~\citep{mesaros2016tut}) task using log-Mel spectrogram representations at a sample rate of 16\,kHz that are extracted in a preprocessing step.
Importantly, tagging and sharing configuration files allows for a one-to-one reproduction of each experiment (assuming that added code is publicly available), as these files determine all the different aspects of the training process -- including random seeds.

\begin{listing}[t]
\small
\begin{minted}{yaml}
defaults:
  - _autrainer_
  - _self_
results_dir: results
experiment_id: default
iterations: 5
hydra:
  sweeper:
    params:
      +seed: 1
      +batch_size: 32
      +learning_rate: 0.001
      dataset: DCASE2016Task1-16k
      model: Cnn10
      optimizer: Adam
\end{minted}
\caption{Exemplary \au configuration file for training a CNN10 model (similar to the model illustrated in \cref{conf:cnn10-random-crop}) on the DCASE2016Task1-16k dataset with log-Mel spectrogram representations extracted using the pipeline outlined in \cref{conf:preprocessing}.}
\label{conf:entry}
\end{listing}

\subsection{Workflow}
\begin{figure}[t]
    \centering
    \includegraphics[width=\columnwidth]{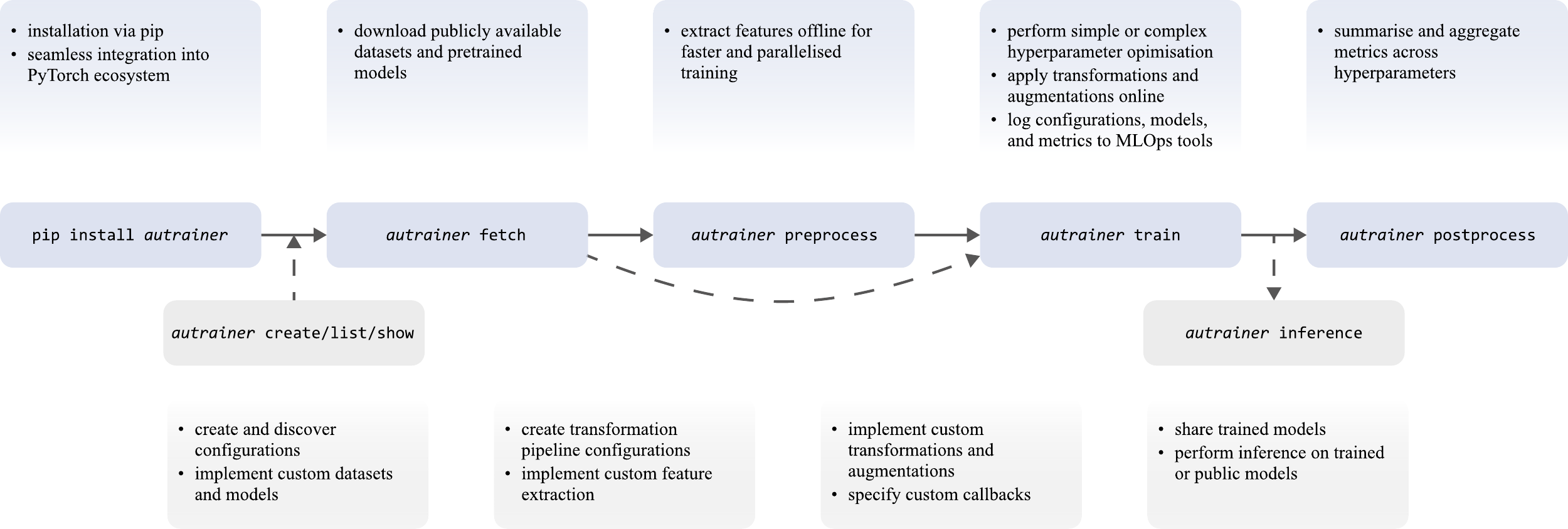}
    \caption{
    Schematic diagram of the \au workflow.
    The package can be installed via \texttt{pip} (or any other \texttt{Python} package manager of choice).
    Subsequently, the user has to specify datasets and models they want to train and a set of possible hyperparameters.
    \texttt{autrainer fetch} can be used to download datasets and model weights, while \texttt{autrainer preprocess} optionally performs offline feature extraction, and \texttt{autrainer train} conducts the training for each set of hyperparameters.
    Finally, \texttt{autrainer postprocess} can be used to summarise and aggregate results.
    The blue cards above the \au commands indicate the key functionality provided by \au while the grey cards below describe optional steps to extend or customise the functionality of the corresponding commands.
    }
    \label{fig:autrainer}
\end{figure}

The overall workflow for \au is shown in \cref{fig:autrainer}.
Our goal is to make the use of the package as easy as possible; thus, we provide a main \ac{CLI} entrypoint which allows the user to get started with model training as quickly as possible (even without writing a single line of code if they wish to use one of the prepackaged datasets).
The choice to split up the main workflow in three steps, namely \texttt{fetch}, \texttt{preprocess}, and \texttt{train} is also made to accommodate for parallel execution of hyperparameter search, e.\,g., allows for parallel training by avoiding race conditions.
An additional \texttt{postprocess} commands allows for an optional summarisation of results.

\subsection{Data pipeline -- \texttt{autrainer fetch}}
The \texttt{fetch} command is responsible for preparing the raw audio data.
This command is responsible for downloading the data by calling the \texttt{autrainer fetch} \ac{CLI} command.
We aim to continually expand the datasets that can be used off-the-shelf -- and invite the community to contribute in this effort -- but the latest version of \au already includes the datasets outlined in \Cref{tab:audio_datasets}.

If the user wishes to work with a dataset which is not included in the public release (e.\,g., because the data itself is not public), they need to write a class that inherits from \texttt{autrainer.datasets.AbstractDataset} and handles the automatic download of the data (if needed) and its transform into a standard format used internally by \au.
This step is only needed if the user wants to implement a new dataset; in case they want to use the original format of datasets already integrated in \au, they can simply proceed with training.

\begin{table}[t]
\small
\centering
\caption{
Overview of Datasets Supported by \au.
Most datasets are publicly available and can be automatically downloaded, while those marked with $^*$ require a request from the original authors.
}
\begin{tabularx}{\textwidth}{p{2cm}lX}
\toprule
\textbf{Task} & \textbf{Dataset} & \textbf{Description} \\
\midrule
\emph{Speech Emotion Recognition}
  & FAU-AIBO$^*$
  & The FAU Aibo Emotion Corpus comprises 18\,216 emotional speech utterances from 51 German children interacting with a robot, recorded at two German schools.
  Each utterance is downsampled to 16\,kHz, labelled at the word level into 11 emotions, and later aggregated into two or five valence classes~\citep{steidl2009automatic}.\\
  & MSP-Podcast$^*$
  & The MSP-Podcast Corpus consists of over 150\,000 emotional utterances extracted from podcast recordings, all sampled at 16\,kHz.
  Each recording is annotated into nine emotion classes and three emotional attributes through crowdsourcing~\citep{lotfian2017building}.\\
  & EmoDB
  & The Berlin Database of Emotional Speech comprises 535 utterances recorded by 10 German actors at at 16\,kHz.
  The dataset includes both short and long utterances which are categorised into seven different emotions~\citep{burkhardt2005database}.\\
\midrule
\emph{Acoustic Scene Classification} 
  & DCASE16-T1
  & The TUT Acoustic Scenes 2016 dataset contains 1511 30-second binaural recordings across 15 acoustic scenes, captured with in-ear microphones at 44.1\,kHz.
  The evaluation set comprises annotations from both expert and non-expert listeners~\citep{mesaros2016tut}.\\
  & DCASE2020-T1A
  & The TAU Urban Acoustic Scenes 2020 dataset comprises 13\,962 10-second training and 2968 validation samples captured across 10 different acoustic scenes.
  The audio samples are recorded with real and simulated mobile devices at 44.1\,kHz~\citep{heittola2020acoustic}.\\
\midrule
\emph{Ecoacoustics} 
  & EDANSA2019
  & The Ecoacoustic Dataset from Arctic North Slope Alaska comprises over 27 hours of audio collected from 40 locations across the Alaskan North Slope.
  The recordings are sampled at 48\,kHz and categorised into four high-level environmental classes~\citep{coban2022edansa}.\\
  & DCASE2018-T3
  & The DCASE2018 Task 3 dataset comprises over 35\,000 10-second audio clips for detecting the presence of bird sounds.
  It combines multiple datasets, including freefield1010~\citep{stowell2013freefield} and BirdVox-DCASE-20k~\citep{lostanlen2018birdvox}, all sampled at 44.1\,kHz~\citep{stowell2018automatic}.\\
\midrule
\emph{Keyword Classification} 
  & SpeechCommands (v2)
  & The Speech Commands dataset consists of over 100,000 one-second utterances of 35 spoken words and background noise.
  Each recording features a single-word command sampled at 16\,kHz~\citep{warden2018speech}. \\
\midrule
\emph{Audio Tagging} 
  & AudioSet
  & The AudioSet dataset contains over two million 10-second audio clips from YouTube, categorised into 527 sound event classes by human annotators.
  All recordings are sampled at 16\,kHz and span a wide range of sounds, including human and animal noises, musical instruments, and everyday environmental sounds~\citep{gemmeke2017audio}.\\
\bottomrule
\end{tabularx}
\label{tab:audio_datasets}
\end{table}

\subsection{Feature extraction -- \texttt{autrainer preprocess}}
\label{sec:feature_extraction}
\au supports a variety of signal transforms for feature extraction, as summarised in \Cref{tab:feature_extraction}.
In addition to feature extraction, \au enables chaining multiple transforms into complex pipelines, offering a high degree of flexibility for constructing complex transform pipelines.
Furthermore, every transform includes an \emph{order} attribute, determining its placement within the pipeline.
This order allows for precise control over the sequence of transforms, enabling specific model requirements to be easily integrated, such as applying normalisation or data augmentation at different stages of the pipeline.

\begin{table}[ht]
\small
\centering
\caption{
Overview of feature extraction and utility transforms supported \au.
}
\begin{tabularx}{\textwidth}{lX}
\toprule
\textbf{Transform} & \textbf{Description} \\
\midrule
\emph{\os features} 
  & \os is our widely-used feature extraction toolkit for speech analysis tasks~\citep{eyben2010opensmile}.
  It bundles numerous feature sets, such as the well-known \textit{eGeMAPS}~\citep{eyben2015geneva} or the official feature set of our INTERSPEECH ComParE Challenge series~\citep{schuller2016interspeech}, and can be extended using configuration files.
  We utilise its \textit{Python wrapper}\footnote{\url{https://audeering.github.io/opensmile-python/}}. \\
\midrule
\emph{\hf transforms} 
  & As several of our supported models are released on \hf, like \wav or \hub, we allow the user to call a \hf \texttt{FeatureExtractor} class which implements the transforms needed for a given model to facilitate the interoperability of \au with the \hf ecosystem. \\
\midrule
\emph{\pann log-Mel spectrograms} 
  & Given the success of \pann models~\citep{kong2020panns}, such as \textit{CNN10} or \textit{CNN14}, we also include their log-Mel spectrogram feature extraction, which relies on the \textit{torchlibrosa} package\footnote{\url{https://github.com/qiuqiangkong/torchlibrosa}}. \\
\midrule
\emph{\ds transforms} 
  & In addition, we offer utility functions that can transform spectrograms (or, in principle, any other 2D feature representation) to an image representation such that models pretrained on computer vision tasks can process them -- i.\,e., \ds models~\citep{amiriparian2017snore}.
  We offer two alternatives: simply upsampling the 2D spectrogram images to a 3D tensor, or converting them to \textit{PNG} images first (as our original \ds work~\footnote{\url{https://github.com/DeepSpectrum/DeepSpectrum}}~\citep{amiriparian2017snore}). \\
\midrule
\emph{Utility} 
  & Finally, we offer a set of utility transforms that can be combined with any of the above methods, including normalisation, random cropping, padding, or replicating the signal to a specified length, i.\,e., covering the most commonly-used transforms in audio processing similar to existing toolkits for feature extraction~\footnote{\url{https://github.com/audeering/audtorch}}. \\
\bottomrule
\end{tabularx}
\label{tab:feature_extraction}
\end{table}

\begin{listing}[t]
\small
\begin{minted}{yaml}
file_handler:
  autrainer.datasets.utils.AudioFileHandler:
    target_sample_rate: 16000
pipeline:
  - autrainer.transforms.StereoToMono
  - autrainer.transforms.PannMel:
      sample_rate: 16000
      window_size: 512
      hop_size: 160
      mel_bins: 64
      fmin: 50
      fmax: 8000
      ref: 1.0
      amin: 1e-10
      top_db: null
\end{minted}
\caption{Preprocessing pipeline extracting mono-channel log-Mel spectrogram representations at a sample rate of 16\,kHz.}
\label{conf:preprocessing}
\end{listing}

Importantly, \au provides the option to apply these transforms both \textit{offline} and \textit{online}, enhancing its adaptability for diverse tasks.
Offline transforms are specified as part of a preprocessing pipeline and are executed once during dataset preparation, via the \texttt{autrainer preprocess} command.
These transforms are included in the dataset configuration and the transformed representation is stored alongside the raw audio files or in a folder designated by the user.
\Cref{conf:preprocessing} illustrates a preprocessing pipeline for extracting mono-channel log-Mel spectrogram representations from audio files sampled at 16\,kHz.
In contrast, online transforms provide greater flexibility by allowing integration into either the model or dataset configurations, allowing for dynamic data transforms during training.
These can be applied globally across all dataset subsets, or customised separately for training, validation, and testing.
\Cref{conf:cnn10-random-crop} illustrates the application of random cropping as an online transform only during training, while leaving the validation and test sets unchanged for consistent evaluation.

\begin{listing}[t]
\small
\begin{minted}{yaml}
id: Cnn10
_target_: autrainer.models.Cnn10
transform:
  type: grayscale
  train:
    - autrainer.transforms.RandomCrop:
        size: 301
        axis: -2
\end{minted}
\caption{Model configuration applying random cropping of input spectrograms for the training subset online.}
\label{conf:cnn10-random-crop}
\end{listing}

\subsubsection{Data augmentation}
\au includes a range of standard data augmentation methods commonly used in computer audition tasks which are summarised in \Cref{tab:data_augmentation}.
Similar to transforms, augmentations have an order attribute to define the order of the augmentations.
The augmentations are combined with the transform pipeline and sorted based on the order of the augmentations as well as the transforms.
In addition to the order of the augmentation, a seeded probability $p$ of applying the augmentation can be specified.
Important: Augmentations from external libraries are not necessarily reproducible, we can only reproduce the probability of applying them but not the actual modification of the input.
To create more complex augmentation pipelines, sequence and choice nodes can be used to create pipelines that resemble graph structures.

\begin{table}[t]
\small
\centering
\caption{
Overview of data augmentations supported by \au.
}
\begin{tabularx}{\textwidth}{lX}
\toprule
\textbf{Augmentation} & \textbf{Description} \\
\midrule
\textit{SpecAugment} 
  & We offer the standard transforms proposed in \textit{SpecAugment}~\citep{park2019specaugment}, namely time masking, frequency masking, and time warping. \\
\midrule
\textit{Gaussian Noise} 
  & We support adding white Gaussian noise to the input signal, simulating real-world noise interference. \\
\midrule
\textit{MixUp and CutMix} 
  & We implement \textit{MixUp}~\citep{zhang2018mixup} and \textit{CutMix}~\citep{yun2019cutmix}, two techniques that interpolate between different signals contained within a batch (and accordingly adjust their labels). \\
\midrule
\textit{External Libraries} 
  & We provide interfaces to external libraries such as \textit{torchaudio}, \textit{audiomentations}, and \textit{torch-audiomentations} for audio processing, as well as \emph{torchvision} and \textit{albumentations} for feature manipulation after transforming audio signals into images. \\
\bottomrule
\end{tabularx}
\label{tab:data_augmentation}
\end{table}

\subsection{Model training -- \texttt{autrainer train}}
Model training is started by calling the \texttt{autrainer train} \ac{CLI} command.
This command utilises the general configuration structure of \au, and allows the user to specify the models and data over which these should be trained, as well as different criterions (i.\,e., loss functions), optimisers, (learning rate) schedulers, and other hyperparameters to search over.
As configuration management is handled by \hydra, \au inherits all hyperparameter optimisation functionality, such as the one supported by \optuna~\cite{optuna2019optuna}.
Moreover, we support all PyTorch optimisers and schedulers.

\subsubsection{Logging}
Building on its internal logging and tracking -- which store model states and outputs -- \au offers interfaces to widely used \ac{MLOps} libraries, such as \mlflow~\citep{zaharia2018accelerating} and \tensorboard~\citep{abadi2015tensorflow}.
Additionally, it provides extensibility for integration with tools like \textit{Weights \& Biases}~\cite{biewald2020experiment}.

\subsubsection{Supported tasks}
\label{sec:supported_tasks}
Currently, \au only supports the tasks of single- and multi-label classification and regression (both single- and multi-target).
For each task, we provide a range of commonly-used losses and metrics, such as the (balanced) cross-entropy loss for classification and mean squared error for regression.
Our long-term goal is to add support for additional tasks, such as \ac{AAC} or \ac{SED}.

\subsubsection{Supported models}
\au includes a constantly-growing list of common models and model architecture families outlined in \Cref{tab:model_architectures} that are used for audio tasks.
These models are configurable by allowing for an adaptation of their standard hyperparameters (length, depth, kernel sizes, etc.).

\begin{table}[ht]
\small
\centering
\caption{
Overview of model architectures supported by \au.
}
\begin{tabularx}{\textwidth}{lX}
\toprule
\textbf{Model} & \textbf{Description} \\
\midrule
\emph{FFNN} 
  & Baseline feed-forward neural networks that can be configured according to the number of hidden layers, width, and other standard parameters.
    These allow the user to train a model using standard, fixed-length features, such as \os functionals. \\
\midrule
\emph{SeqFFNN} 
  & An extension of the above, sequence-based FFNNs, which first process dynamic features with a sequential model, such as Long short-term memory (LSTM)~\citep{hochreiter1997long} or Gated Recurrent Unit (GRU)~\citep{chung2014empirical} networks. \\
\midrule
\emph{CRNN} 
  & End-to-end, convolution-recurrent neural networks (CRNNs)~\citep{tzirakis2018end, zhao2019speech} adapted from our End2You toolkit~\footnote{\url{https://github.com/end2you/end2you}}. \\
\midrule
\emph{PANN} 
  & The two best-performing models from PANNs, namely CNN10 and CNN14~\citep{kong2020panns}.
    These models can be both trained from scratch or fine-tuned from the weights released by the original authors. \\
\midrule
\emph{TDNNFFNN} 
  & The Time-Delay Neural Network (TDNN)~\citep{desplanques2020ecapa} pretrained on VoxCeleb1~\citep{nagrani2020voxceleb} \& VoxCeleb2~\citep{chung2018voxceleb2} included in SpeechBrain~\citep{ravanelli2021speechbrain}\footnote{\url{https://huggingface.co/speechbrain/spkrec-ecapa-voxceleb}} as a backbone to extract embeddings, which are then passed to a configurable FFNN for the final prediction. \\
\midrule
\emph{ASTModel} 
  & The Audio Spectrogram Transformer (AST), optionally pretrained on AudioSet~\citep{gong2021ast}. \\
\midrule
\emph{LEAFNet} 
  & LEAFNet incorporates LEAF (Learnable Efficient Audio Frontend) and the additional components, as implemented either in the original work~\citep{zeghidour2021leaf} and included in SpeechBrain or the follow-up work of \citet{meng2023what}. \\
\midrule
\emph{W2V2FFNN} 
  & \wav~\citep{baevski2020wav2vec} and \hub~\citep{hsu2021hubert} models to extract audio embeddings, followed by a configurable FFNN as in \citet{wagner2023dawn}. We support all different \hf variants of \wav and \hub. \\
\midrule
\emph{WhisperFFNN} 
  & Similar to the above, but using Whisper instead of \wav or \hub~\citep{radford2022robust}. \\
\midrule
\emph{DeepSpectrum} 
  & Similar to \ds~\citep{amiriparian2017snore}, we allow the processing of spectrograms using image-based models, and add support for all the ones included in Torchvision~\citep{torch2016torchvision} and Timm~\citep{wrightman2019timm}, both with randomly-initialised weights and their pretrained versions. \\
\bottomrule
\end{tabularx}
\label{tab:model_architectures}
\end{table}

\subsection{Postprocessing interface -- \texttt{autrainer postprocessing}}
Beyond the core training functionality, \au can process any finished training pipeline in an optional, customisable and extensible postprocessing routine acting on the saved training logs.
This offers particular usability for grid searches over large hyperparameter spaces, summarising training curves and model performances across runs.
\au further allows for the aggregation of trainings across certain (sets of) hyperparameters, such as random seeds or optimisers, in terms of average performance.

\subsection{Inference interface -- \texttt{autrainer inference}}
\au includes an inference interface, which allows to use publicly-available model checkpoints and extract both (sliding-window-based) model predictions and embeddings from the penultimate layer.
This can be done with the \texttt{autrainer inference} \ac{CLI} command.
As part of the official release, we additionally provide pretrained models on Hugging Face~\footnote{\url{https://huggingface.co/autrainer}} for speech emotion recognition, ecoacoustics, and acoustic scene classification.
We offer detailed model cards and usage instructions for each published model.

\section{\au design principles}
In the previous sections, we have described the key features of \au.
In the present section, we reiterate our key design considerations and highlight the strengths of our package.

%\subsection*{Reproducibility}
A major emphasis of our work was placed on the reproducibility of machine learning experiments for computer audition.
This has been ensured by the consistent setting of random seeds, and the strict definition of all experiment parameters in configuration files.
While we do not take any steps to ensure that these configuration files cannot be tampered with, our workflow nevertheless enables researchers to reproduce the work of original authors given the latter have released their configuration files and the corresponding \au version.

%\subsection*{Baselines}
\au allows a fair comparison with a number of readily-available `standard' baselines for each dataset.
Specifically, a user can rely on its grid-search functionality to compare their new model architecture to baseline models using the same hyperparameters and computational budget.
This reduces the considerable workload of having to implement existing baselines from scratch (e.\,g., by porting code from non-maintained repositories) and should help with the comparability of different methods.

%\subsection*{Low-code training}
\au lowers the barrier of entry to the field of computer audition.
For example, in the case of computational bioacoustics, several of the expected users are biologists with little training in machine learning applications.
Relying on \au for the machine learning aspects allows them to benefit from advances in that field, while only caring for implementing a dataset class that iterates through their data.

Table~\ref{tab:comparison} provides a comparative overview of \au and related audio \ac{DL} toolkits.

\begin{table}[t]
\small
\centering
\caption{Comparison of audio \ac{DL} toolkits in terms of feature extraction, model training, and experiment management capabilities.}
\begin{tabularx}{\textwidth}{lXXX}
\toprule
\textbf{Toolkit} & \textbf{Feature Extraction} & \textbf{Model Training} & \textbf{Experiment Management} \\
\midrule
openSMILE & Hand-crafted acoustic descriptors & Not provided & Not provided \\
Librosa & log-Mel spectrograms, \acp{MFCC} & Not provided & Not provided \\
Audiomentations & Waveform and spectrogram augmentations & Not provided & Not provided \\
auDeep & Unsupervised spectral embeddings & \acp{SVM}, \acp{MLP} & \ac{CLI} \\
DeepSpectrum(Lite) & Spectrogram features and augmentations & \acp{CNN} & TOML and JSON configurations \\
End2You & End-to-end audio and spectrogram & \acp{CNN}, \acp{RNN} & CLI \\
nkululeko & Comprehensive feature extraction and augmentations & Classical \ac{ML} and \ac{DL} & ini-file configuration \\
SpeechBrain & Feature extraction, waveform and spectrogram augmentations & \ac{LM} focus & YAML configuration \\
ESPNet & Feature extraction, waveform and spectrogram augmentations & \acp{CNN}, \acp{RNN}, Transformers & YAML configuration \\
\midrule
\au & Pipeline-based feature extraction, waveform, and spectrogram augmentations & \acp{MLP}, \acp{CNN}, \acp{RNN}, Transformers & YAML configuration \\
\bottomrule
\end{tabularx}
\label{tab:comparison}
\end{table}

\section{Results}
To validate the applicability of \au, we train several models across common computer audition tasks.
Experimental results are summarised in Table~\ref{tab:results}, which details each task, dataset, model architecture, utilised features, and achieved performance.
The trained model checkpoints, along with detailed descriptions, are publicly available on Hugging Face\footnote{\url{https://huggingface.co/autrainer}}.

\begin{table}[t]
\small
\centering
\caption{Experimental results obtained using \au.}
    \begin{tabularx}{\textwidth}{lXXXX}
    \toprule
    \textbf{Task} & \textbf{Dataset} & \textbf{Model} & \textbf{Features} &     \textbf{Performance} \\
    \midrule
    Acoustic Scene Classification & DCASE2020-T1A & CNN14 & log-Mel & $.678$ accuracy \\
    Ecoacoustics & EDANSA2019 & CNN10 & log-Mel & $.871$ weighted F1 \\
    Speech Emotion Recognition & MSP-Podcast & Wav2Vec2-Large-12 & raw audio & $.650$ unweighted average recall\\
    \bottomrule
    \end{tabularx}
\label{tab:results}
\end{table}

\section{Future roadmap}
By publicly releasing \au, we wish to engage with the larger audio community to further expand the capabilities of our toolkit.
Our goal is to expand our offering of off-the-shelf datasets to include the most commonly used benchmarks and domain-specific datasets across different computer audition tasks.
Currently, \au only supports standard classification, regression, and tagging.
In the future, we aim to expand it for \ac{AAC}, \ac{SED}, and \ac{ASR} by incorporating the appropriate losses and data pipelines.
We will additionally incorporate both specific model architectures and fundamentally different classes of models -- such as large audio models~\citep{triantafyllopoulos2024computer} -- in juxtaposition with the tasks and datasets that will be added.

\section{Conclusion}
\label{sec:conclusion}
This work described \au, an open-source toolkit aimed at computer audition projects that rely on deep learning.
We have outlined all major features and design principles for the current version of \au.
Our main goals were to offer an easy-to-use, reproducible toolkit that can be easily configured and used as a low- or even no-code option.
We look forward to a more engaged conversation with the wider community as we continue to develop our toolkit in the years to come.

\section*{Acknowledgements}
This work has received funding from the DFG's Reinhart Koselleck project No.\ 442218748 (AUDI0NOMOUS), the DFG project No.\ 512414116 (HearTheSpecies), and the EU H2020 project No.\ 101135556 (INDUX-R).
We additionally thank our colleague, Alexander Gebhard, for being an early adopter of our toolkit and delivering useful feedback during the early development phase.

\section{\refname}
\printbibliography[heading=none] 
\end{document}